\documentclass[12pt]{article}

\usepackage{latexsym,amsmath,amssymb,pstricks,wasysym,stmaryrd,enumerate,epsfig}

\newcommand{\be}{\begin{equation}}
\newcommand{\ee}{\end{equation}}
\newcommand{\beu}{\begin{equation*}}
\newcommand{\eeu}{\end{equation*}}
\newcommand{\bea}{\begin{eqnarray}}
\newcommand{\eea}{\end{eqnarray}}
\newcommand{\beaa}{\begin{eqnarray*}}
\newcommand{\eeaa}{\end{eqnarray*}}
\newcommand{\bmx}{\begin{pmatrix}}
\newcommand{\emx}{\end{pmatrix}}

\newcommand{\del}{\partial}
\newcommand{\g}{{\mathfrak g}}
\newcommand{\h}{{\mathfrak h}}
\newcommand{\m}{{\mathfrak m}}

\newcommand{\hs}{{*}}

\newcommand{\half}{\frac{1}{2}}

\newcommand{\nn}{\nonumber}

\newcommand{\8}{{\infty}}

\newcommand{\eps}{\epsilon}
\newcommand{\tr}{\,{\rm tr}\,}

\newcommand{\hc}{{\dagger}}
\newcommand{\qbar}{{\bar q}}

\advance\textheight by \topskip
\begin{document}

\baselineskip 17pt
\parindent 8pt
\parskip 9pt

\begin{flushright}
\break
hep-th/0503008\\[3mm]
\end{flushright}
\vspace{1cm}
\begin{center}
{\Large {\bf Non-local charges, $\mathbb Z_m$ gradings}}

{\Large {\bf and coset space actions}}\\[4mm]
\vspace{0.8cm} 
{\large C. A. S. Young,}
\\
{\em Department of Mathematics, University of York,\\
Heslington Lane, York YO10 5DD, UK}
\\
{\small E-mail: {\tt charlesyoung@cantab.net}}
\\

\end{center}

\vskip 0.15in
 \centerline{\small\bf ABSTRACT}
\centerline{
\parbox[t]{5in}{\small
\noindent
The construction of flat currents, and hence conserved non-local charges, for the
superstring on $AdS_5\times S^5$ is generalised. It is shown that such currents exist
for sigma-model type actions on all coset (super-)spaces $G/H$ in which, at the level of 
the Lie algebras, $\h$ is the grade-zero subspace of a $\mathbb Z_m$-grading of $\g$.  
This is true for an essentially unique choice of the Wess-Zumino term, which is determined. 
}}

\vspace{1cm}

\section{Introduction}
There has been much recent interest in the role of integrability in the world-sheet theory of type IIB strings in $AdS_5\times S^5$. 
In \cite{BPR}, Bena, Polchinski and Roiban found an infinite number of non-local classically conserved charges for the 
Green-Schwarz superstring in this background. Subsequently Vallilo showed \cite{Vallilo:2003nx} that such charges also exist in
the pure-spinor formalism for the superstring. These charges are the analogues of the non-local charges which have long been known
to exist in the sigma models on symmetric spaces \cite{EF,EFlocal,Schwarz:1995td,Mandal:2002fs}, and their 
discovery allowed ideas 
from integrable field theory to be applied to the world-sheet theory of superstrings on $AdS_5\times S^5$ 
\cite{Hatsuda:2004it, Das:2004hy,Alday:2005gi}. In the pure-spinor formalism it has been argued that the charges survive 
quantum-mechanically \cite{Berkovits:2004jw,Berkovits:2004xu}. 
In the context of the AdS/CFT correspondence\cite{AdSCFT}, which was the initial motivation for the search for these charges, it was 
subsequently shown by Dolan, Nappi and Witten \cite{DNW} that the same Yangian symmetry algebra is present in the weakly coupled 
limit of SYM on the gauge theory side. (Further connections with supersymmetric gauge theory are made in 
\cite{Bksz, Wolf:2004hp,Agarwal:2004sz}.) 

The charges are constructed by first identifying a family of currents $a(\mu)$, depending 
smoothly on a spectral parameter $\mu$, that are valued in some Lie-algebra and that are flat:
\be da(\mu)+ a(\mu)\wedge a(\mu) =0.\ee
One then constructs the monodromy matrix
\be T_{(\mu)}(t) = P\exp \int_{(-\infty,t)}^{(+\infty,t)} a(\mu) \label{monodromy}\ee
which is conserved by virtue of the flatness of $a$, and the non-local charges are obtained by expanding $T_{(\mu)}$ in powers of the
spectral parameter (for the details see e.g. \cite{BPR,MacKay:2004tc}). 
The crucial step in \cite{BPR} was thus the identification of the family of flat currents $a(\mu)$. 
The fact that this was possible appeared to be something of a coincidence. 

In this paper we put these currents in a more general context, with the hope that this will eventually allow a deeper 
understanding of why they exist at all, and what role they play. Let us first recall what it is about
the target space $AdS_5\times S^5$ that allows their construction.

The space $AdS_5\times S^5$ is the bosonic part of the coset superspace
\be \frac{PSU(2,2|4)}{SO(1,4)\times SO(5)}.\label{css}\ee
and the Green-Schwarz superstring action can be thought of as a sigma model-type action on this space 
\cite{Metsaev:1998it} (see also \cite{Roiban:2000yy}, in which the coset space is chosen slightly differently).
This space is not a symmetric space but it does have similar properties. Recall that a coset
space $G/H$ is said to be symmetric if $H$ is the fixed point set of an automorphism $\sigma$ of $G$ of order 2. That is, at the level 
of the algebras,
\be \sigma:\g \rightarrow \g,\quad \sigma[X,Y] = [\sigma X,\sigma Y], \quad \sigma^2 = 1\ee
so that the decomposition
\be \g = \h + \m \ee
into the $(+1)$- and $(-1)$-eigenspaces of $\sigma$ is a $\mathbb Z_2$ grading of $\g$:
\be [\h,\h] \subset \h,\quad [\h,\m] \subset \m, \quad [\m,\m] \subset \h.\ee
The existence of this automorphism $\sigma$, and the resulting $\mathbb Z_2$ grading, turn out to be crucial in the construction of 
flat currents in bosonic sigma models. In the coset superspace (\ref{css}) the subgroup $H$ is again the fixed point set of an 
automorphism of $G$, but this automorphism is now of order 4 \cite{BBHZZ}. There is thus a $\mathbb Z_4$ grading of $\g$ and 
it is this grading which allows the construction of the flat currents. 

As was noted in \cite{Vallilo:2003nx}, this means that the same construction applies equally well to other coset superspaces 
with a $\mathbb Z_4$ grading, including for example
\be \frac{PSU(1,1|2)\times PSU(2|2)}{SU(2)\times SU(2)}\quad \text{and}\quad \frac{PSU(1,1|2)}{U(1)\times U(1)} \ee
whose bosonic parts are $AdS_3\times S^3$ and $AdS_2\times S^2$ respectively \cite{BBHZZ}. 

But in fact nothing in the construction even relies on $G$ being a supergroup. As we shall discuss, it is possible to take \emph{any} 
Lie group $G$ whose algebra $\g$ admits a $\mathbb Z_4$-grading and construct actions on the coset space $G/H$, where $H$ is the 
subgroup  corresponding to the grade-zero subalgebra $\h$. Provided the WZ term is correctly 
normalised, the resulting theories possess non-local charges.

One natural question this raises is: if there are non-local charges for theories on the coset spaces associated with $\mathbb Z_2$ 
gradings and $\mathbb Z_4$ gradings, what about gradings of arbitrary finite order $m$? In fact, before one even addresses the issue 
of non-local charges, the question of what actions \emph{exist} for fields in such coset spaces is interesting in its own right. 
For, as we recall below, even when $m=4$ there are two natural choices: the Green-Schwarz-type action, which has kinetic 
terms only for the target-space bosons, and the ``hybrid'' action (as in \cite{BBHZZ}) which has kinetic terms for both the target-space 
bosons and fermions and which is used in the pure-spinor description of the superstring. 

This paper thus has two aims: first, to construct actions on general coset spaces defined by gradings of $\g$ of finite order, 
and second to identify actions that possess non-local symmetries. For this latter step we shall restrict our attention to the 
simplest (sigma-model, or ``hybrid model''-type) kinetic term, and will find that there are flat currents and so non-local symmetries 
for a suitable choice of the Wess-Zumino term. 

In section \ref{Z3} we deal with the simplest new case, that of coset spaces $G/H$ defined by third order automorphisms. 
Then in section \ref{Zm} we generalise the discussion to automorphisms of arbitrary finite order.

\subsection{Notation}

We begin by fixing some notation and assumptions. In what follows the worldsheet coordinates are $(t,x)=(x^{0},x^{1})$ and worldsheet 
vector indices are drawn from $\mu, \nu, \rho, \dots$. The worldsheet metric and alternating symbol are 
\be \eta_{\mu \nu} = \bmx 1 & 0 \\ 0 & -1 \emx, \quad \eps_{\mu \nu} = \bmx 0 & 1 \\ -1 & 0 \emx.\ee
The identities $\hs \hs a= +a$ and $a\wedge \hs b + \hs a \wedge b =0$ for differential one-forms $a,b$ are used frequently. 

Let $g(t,x)$ be a field valued in a faithful matrix representation of a (super-)group $G$. 
Currents, like $g^{-1} dg$, are valued in the corresponding
matrix representation of the Lie (super-)algebra $\g$. Let \be\tr(X)\ee 
denote the trace (or the appropriate supertrace) of $X$ in our chosen representation.\footnote{\label{str}In what follows we will 
sometimes refer simply to the ``trace'', with the understanding that this is a supertrace when $\g$ is a superalgebra.}

We assume that $\g$ is $\mathbb Z_m$-graded. That is, we assume there is a decomposition
\be \g = \sum_{k=0}^{m-1} \g_{(k)} \label{realgrad}\ee
(here $\g_{(0)}=\h$ is the Lie algebra of $H$) that respects the Lie bracket:
\be \left[ \g_{(i)} , \g_{(j)} \right] \subset \g_{(i+j)} \ee
where the addition of the indices is understood to be modulo $m$. Further, we assume that the trace is compatible with the 
grading, in the sense that if $X\in \g_{(i)}$ and $Y\in \g_{(j)}$ then
\be \tr XY = 0  \quad\text{unless}\quad i+j\equiv 0 \mod m.\label{gradd} \ee 

The other properties of the (super)trace we shall require are cyclicity\be\tr WX\dots YZ = \tr ZWX\dots Y\ee 
(which in particular implies ${\rm Ad}(G)$-invariance of the inner product: $\tr XY = \tr UXU^{-1} UYU^{-1}$ for all $U\in G$)
and non-degeneracy, in the sense that if $Y\in \g_{(i)}$ then \be\tr XY = 0 \;\;\forall X\in \g_{(m-i)} \implies Y=0.\ee

\subsection{Gradings, automorphisms and a family of examples}

It is worth commenting briefly on the construction of $\mathbb Z_m$ gradings. The $\mathbb Z_m$ gradings 
of a \emph{complex} Lie algebra $\g^\mathbb C$ correspond to automorphisms of $\g^\mathbb C$ of 
order $m$, and for the simple Lie algebras these automorphisms are known and classified (see \cite{Kac:1990gs,Goddard:1986bp} for
the bosonic cases and \cite{Serganova,S2} for the supersymmetric cases). However, not every real form $\g$ of $\g^\mathbb C$ will be
compatible with a given grading.

An example makes this clear: let $a,b,c$ be positive integers and consider the $\mathbb Z_3$ grading of 
$\mathfrak a_{a+b+c-1}=\mathfrak{sl}(a+b+c,\mathbb C)$ defined by the 
automorphism (in the defining representation)
\be \sigma: X \mapsto N X N^\hc \quad \text{where}\quad N = \bmx e^{2\pi i/3}\,1_{a\times a} &&\\ & 1_{b \times b} &\\ && 
e^{-2\pi i/3} \,1_{c\times c}\emx.\ee
The subspaces of grades $0,1,2$ are the eigenspaces of $\sigma$ with eigenvalues $1,e^{2\pi i/3}, e^{-2\pi i/3}$ respectively, 
and consist of the matrices of the block form 
\be \bmx A&&\\ & B &\\ &&C \emx\in \h,\quad \bmx & D & \\ && E\\ F && \emx\in \g_{(1)},\quad \bmx&&G\\H&&\\& K &\emx\in\g_{(2)},\ee 
where  $\tr A +\tr B + \tr C=0$.

For real $A,B,\dots,G,H,K$ these are subspaces of the real form $\mathfrak{sl}(a+b+c,\mathbb R)$, which therefore does admit a 
$\mathbb Z_3$ grading. But they are clearly not subspaces of the \emph{compact} real form $\mathfrak{su}(a+b+c)$.

Obviously, by redefining $N$ to have the all $m$-th roots of unity down the diagonal, one can construct $\mathbb Z_m$ gradings of 
$SL(a_1+a_2+\dots+a_m,\mathbb R)$ in exactly the same way. These coset spaces,
\be\frac{SL(a_1+a_2+\dots+a_m,\mathbb R)}{S\left(L(a_1,\mathbb R)\times L(a_2,\mathbb R) \times \dots \times L(a_m,\mathbb R)\right)},\ee
provide one concrete class of examples for the results that follow.

\section{$\mathbb{Z}_3$ gradings and sigma model actions}
\label{Z3}
We consider first models constructed using $\mathbb Z_3$ gradings. Let us write the decomposition of $\g$, and of the current 
$j=g^{-1}dg\in \g$, as 
\be \g = \h + \g_{(1)} + \g_{(2)},\quad  j = A + q + \qbar.\ee

We are interested in models whose physical degrees of freedom take values in the space of cosets $\left\{gH:g\in G\right\}=G/H$, 
so we look for actions in which 
\be g\mapsto Ug, U\in G\ee is a global symmetry, while 
\be g\mapsto gh(t,x), h\in H\ee is a local symmetry. Under the former, $j$ is invariant, while under the latter, 
\be j \mapsto h^{-1} j h + h^{-1} dh \label{jtrans}\ee
so that $A$ transforms like a gauge connection while $q,\qbar$ are covariant.

Given the grading property (\ref{gradd}), the only kinetic term one can write down\footnote{This is at least the only possibility 
constructed from the symmetric second-rank tensor $\tr(XY)$ on $\g$ (which, it should be noted, is not in general negative definite). 
It is the invariance of this tensor under the adjoint action of $H$ which guarantees that the action has 
local $H$-symmetry. In many particular cases $(G,H,\sigma)$ there are other independent tensors on $\g$ with this property. 
Similarly, to construct WZ terms one requires antisymmetric third-rank tensors invariant under $Ad_H$, and there may be others besides 
$\tr [X,Y]Z$. (The cases with $G$ simple are discussed in \cite{Evans:2000qx}, for the symmetric tensors, 
and \cite{Evans:2005zd}, for the antisymmetric tensors, and references therein.) For simplicity we shall consider here only the 
invariants constructed from $\tr()$, which are generic.} with the correct symmetries is
\be -\frac{1}{\lambda^2} \tr q_\mu \qbar^\mu = -\frac{1}{\lambda^2}\tr q \wedge \hs \qbar .\ee
(Here, and throughout, $\lambda$ is some overall coupling which provides a scale for the model. It will not be important since
we are concerned only with the classical dynamics.) 
This is nothing but the usual sigma-model kinetic term on $G/H$ (see e.g. \cite{Evans:2000qx}) since it may be re-written
\be -\frac{1}{2\lambda^2} \tr \left(j-A\right)_\mu \left(j-A\right)^\mu= -\frac{1}{2\lambda^2} \tr (j-A)\wedge \hs (j-A).\ee

In addition to the kinetic term, the symmetries may also allow Wess-Zumino terms -- that is, terms of the form 
$\int_B \omega$, where $B$ is a 3-manifold whose boundary is the worldsheet and $\omega$ is a closed 3-form \cite{Witten:1983ar}.
We must thus find all closed 3-forms with the correct symmetries. There are only two linearly 
independent 3-forms constructed using $\tr$, given the $\mathbb Z_3$ grading, and only one closed linear combination of these, which is
\be \tr \left( q\wedge q\wedge q - \qbar\wedge\qbar\wedge\qbar\right).\ee
This is in fact also exact: it is
\be d \tr q \wedge \qbar.\ee

One computes these exterior derivatives by first noting that the zero curvature identity for $j$,
\be dj +j\wedge j =0,\ee
implies, grade by grade, the identities
\be F + q\wedge \qbar + \qbar \wedge q =0 \label{DA}\ee
\be Dq + \qbar \wedge \qbar = 0 \label{Dq}\ee
\be D\qbar + q\wedge q =0 \label{Dqbar}\ee
where
\be D\omega = d\omega + A\wedge \omega - (-)^{|\omega|} \omega\wedge A \ee
for a Lie algebra valued form $\omega$ of degree $|\omega|$, and
\be F = dA + A\wedge A.\ee
Then by invariance of the trace one has for example
\be d \tr q \wedge q \wedge q = D\tr q \wedge q\wedge q = 
\tr \left(Dq \wedge q \wedge q- q \wedge Dq \wedge q + q \wedge q \wedge Dq\right). \ee

Thus the most general action with the required symmetries is actually local. (This is also true in the $\mathbb Z_4$ 
case \cite{BBHZZ}.) We have
\be S = -\frac{1}{\lambda^2}\int d^2x \tr \left( q_\mu \qbar^\mu + \frac{\gamma}{3}\eps^{\mu \nu} q_\mu \qbar_\nu \right)
= -\frac{1}{\lambda^2}\int \tr \left(  q\wedge \hs \qbar + \frac{\gamma}{3} q \wedge \qbar \right),\ee
where we leave coefficient of the WZ term arbitrary for the moment.

The resulting equations of motion are
\be D_\mu q^\mu - \gamma \eps^{\mu \nu} D_\mu q_\nu = 0 \ee
\be D_\mu \qbar^\mu + \gamma \eps^{\mu \nu} D_\mu \qbar_\nu = 0 \ee
or, in form notation,
\be D\hs q - \gamma Dq = 0\label{qeom}\ee
\be D\hs\qbar + \gamma D\qbar =0,\label{qbareom}.\ee 
These are most conveniently derived by considering variations of the action of the form $g\mapsto g(1+ X)$, so that
\be j\mapsto j + dX + [j,X]= j+ DX + [q+\qbar,X]\ee
and if $X$ has grade 2 under $\sigma$ then
\be \delta q = [\qbar ,X ] \ee
\be \delta \qbar = D X \ee
and we find the equation of motion (\ref{qeom}) for $q$. 
The equation of motion for $\qbar$ is obtained by considering $X$ of grade 1. 

The complete set of equations of motion can be written as the conservation law $d\hs J=0$ for the Noether current associated with 
the left-$G$ symmetry $G\mapsto (1+X)G$, which is
\be J = g\left( q + \qbar -\frac{\gamma}{3} \hs( q - \qbar) \right) g^{-1}.\label{Noether}\ee  

\subsection{Flat currents}
We now seek flat currents, and, we hope, a one-parameter family of them. There is a choice: we can try to construct flat currents 
\emph{invariant} under the gauge symmetry $g\mapsto gh$, or flat currents that transform in the same way (\ref{jtrans}) as $j=g^{-1} dg$.
But the two are essentially equivalent, for suppose $\tilde \jmath\,$ is any flat current of the latter type, 
and write the flatness property as the vanishing of the curvature of the corresponding covariant derivative: 
\be \eps^{\mu\nu}[\del_\mu + \tilde\jmath_\mu ,\del_\mu +\tilde\jmath_\nu] =0;\ee
then clearly the derivative operator $g(\del_\mu +\tilde\jmath_\mu)g^{-1}$ also has vanishing curvature, and, by construction, it is 
gauge-independent. Now 
\be g(\del_\mu + \tilde\jmath_\mu)g^{-1} = \del_\mu + g\left( \tilde\jmath_\mu - j_\mu \right) g^{-1} \label{gaugeinvcurrent}\ee
and so the current $g\left( \tilde\jmath - j \right) g^{-1}$ is flat and gauge-invariant. 
It is straightforward to verify that this is the relationship between the flat currents of Das, Maharana, Melikyan and Sato 
\cite{Das:2004hy} and those originally constructed by Bena, Polchinski and Roiban \cite{BPR}.

Since it is easiest to work with objects having definite grade, we shall look for flat gauge-\emph{dependent} 
currents. These must be of the form
\be j{(\mu)} = A + e(\mu) q + \bar e(\mu) \qbar + f(\mu) \hs q + \bar f(\mu) \hs \qbar \ee 
where $\mu$ is some parameter and $e,\bar e,f,\bar f$ are functions to be determined (and the coefficient of $A$ must clearly
be unity). The curvature of this current is
\bea dj{(\mu)} + j{(\mu)} \wedge j{(\mu)} &=& F + e Dq + \bar e D\qbar + f D\hs q + \bar f D\hs \qbar 
\\&&+ (e^2-f^2) q\wedge q + (\bar e^2-\bar f^2) \qbar \wedge \qbar + \nn\\ &&(e\bar f-\bar e f)( q\wedge\hs\qbar - \qbar\wedge\hs q)
+(e\bar e- f\bar f)(q\wedge \qbar + \qbar \wedge q).\nn\eea  
On making use of the equations of motion (\ref{qeom} and \ref{qbareom}) and the zero curvature identity for $j$ 
(\ref{DA}, \ref{Dq}, \ref{Dqbar}), one finds that this vanishes provided 
\be e\bar e - f\bar f = 1, \quad e\bar f -\bar e f =0\ee\be e + \gamma f = \bar e^2 - \bar f^2,\quad \bar e - \gamma \bar f = e^2-f^2.\ee

The first two of these may be re-written as
\be (e+f)(\bar e - \bar f) = (e-f)(\bar e + \bar f) =1 \ee
and therefore, using now all the equations,
\be 1= (e^2 - f^2)(\bar e^2-\bar f^2) = (e+\gamma f)(\bar e-\gamma \bar f).\ee
Thus there are extra solutions, in addition to the current $j=g^{-1}dg$ ($e=\bar e=1, f=\bar f =0$), only when $\gamma=\pm 1$.
Henceforth we shall take $\gamma = +1$. (There is no loss of generality in this, because reversing the sign of $\gamma$ is 
equivalent to replacing the defining automorphism $\sigma$ with $\sigma^2$.) 

Let us express all the coefficients as functions of the parameter $\mu$ according to 
\be \mu=e+f= (\bar e - \bar f)^{-1} = (\bar e + \bar f)^\half = (e-f)^{-\half} \label{para3}\ee
so that
\bea  e = \frac{\mu^3 +1}{2\mu^2}&\quad& f= \frac{\mu^3-1}{2\mu^2}\\ 
 \bar e = \frac{\mu^3 +1}{2\mu}  &\quad&\bar f = \frac{\mu^3-1}{2\mu}.\eea

\subsection{The monodromy matrix and conserved charges}
The current $j=g^{-1} dg$ corresponds to $\mu=1$. Let us perturb around this by setting $\mu = \exp \theta$, so that, to second order in 
$\theta$,
\bea e = 1- \half \theta+\frac{5}{4}\theta^2 + \dots&\quad& f = \frac{3}{2} \theta -\frac{3}{4} \theta^2 + \dots\\ 
\bar e = 1 + \half \theta +\frac{5}{4} \theta^2 + \dots&\quad& \bar f = \frac{3}{2}\theta +\frac{3}{4}\theta^2+\dots.\eea
The physical, gauge-\emph{invariant} flat current, defined as in (\ref{gaugeinvcurrent}), is thus
\bea g\left( j(\exp \theta) - j \right) g^{-1} &=& g\left( (e-1)q + (\bar e -1)\qbar + f\hs q +\bar f \hs \qbar \right) g^{-1} \nn\\
                               &=& \frac{3}{2} \theta \,\hs g\left( q + \qbar - \frac{1}{3} \hs (q - \qbar) \right) g^{-1} \nn \\
                               && {}+ \frac{3}{4} \theta^2 \,\hs g\left( -q + \qbar + \frac{5}{3} \hs (q+\qbar) \right) g^{-1}+\dots\nn\\
                               &=:& \frac{3}{2} \theta \, \hs J + \frac{3}{4} \, \hs \tilde J+\dots.\label{pgi}\eea
Note that the Noether current (\ref{Noether}), or rather a multiple of its hodge dual, appears here at first order in $\theta$. 
This is as expected, since it means that in the expansion of the monodromy matrix 
\be T_{(\theta)} = P\exp \int_{x=-\8}^{x=+\8} g \left( j(1+\theta)-j\right) g^{-1}\ee
(as defined in (\ref{monodromy})) the usual local Noether charge appears at first order in $\theta$, with a numerical prefactor:
\be Q^{(0)}(t) = -\frac{3}{2} \int_{-\8}^{+\8} dx J_0(t,x).\ee
 
The non-local charges are obtained by expanding to higher orders, but there is some freedom in how this expansion is 
performed and hence in the definitions of these charges. We have already made one choice 
in setting $\mu=\exp \theta$ rather than, say, simply $\mu = 1+ \theta$.\footnote{The virtue of the choice 
$\mu \propto \exp \theta$ is that it preserves the symmetry between $q$ and $\qbar$ in the expansion (\ref{pgi}).
Other choices must still yield conserved charges of course, and indeed setting \be\mu = 1 + \theta + (\half+\alpha)
\theta^2+\dots\ee has the effect of sending \be Q^{(1)} \mapsto Q^{(1)} + \alpha Q^{(0)}.\ee}
There is also the choice of how to expand the monodromy matrix. Let us choose to set
\be T_{(\theta)} =: \exp \left( \theta Q^{(0)} + \theta^2 Q^{(1)} + \dots \right).\ee
With these conventions one finds that the first non-local charge is
\be Q^{(1)}(t) = \int_{-\8}^{+\8} dx 
        \left( -\frac{3}{4} \tilde J_0(t,x) + \frac{9}{8}\int_{-\8}^x dy \left[ J_0(t,x), J_0(t,y)\right] \right)\ee
where $\tilde J$ is defined in (\ref{pgi}).\footnote{As a check, it may be verified directly that this charge is conserved using the 
equations of motion (\ref{qeom},\ref{qbareom}), together with the identity $d(gXg^{-1}) = g(DX + [q,X] + [\qbar,X])g^{-1}$.}

\section{The construction for higher order automorphisms}\label{Zm}
We now generalise the discussion to $\mathbb Z_m$ gradings, for $m>3$. We shall consider in parallel the cases in which $m=2n+1$ 
is odd and in which $m=2n$ is even. There are a few differences but the 
bulk of the argument is the same. (In both cases, $n$ will always denote the greatest integer $\leq m/2$.)

Let us write the decomposition of $\g$ and of the current $j=g^{-1}dg$, as 
\be \g = \h + \sum_{k=1}^{m-1} \g_{(k)},\quad \label{decompj} j = A + \sum_{k=1}^{m-1} q_{(k)}.\ee 

The flatness property of $j$ now implies that
\be F + \sum_{i=1}^{m-1} q_{(i)} \wedge q_{(m-i)} = 0\label{Fid}\ee
and that for every $k \in \{1,2,\dots,m-1\}$
\be Dq_{(k)} + \sum_{(i,j) \in (\mathbb Z_m \setminus \{0\})^2}^{i+j\equiv k} q_{(i)} \wedge q_{(j)} =0,\label{Dqk}\ee
where $\equiv$ denotes congruence modulo $m$. (The apparent double counting in the summations is deliberate: for example in (\ref{Dqk})
whenever $q_{(i)}$ and $q_{(j)}$ are distinct, both $q_{(i)}\wedge q_{(j)}$ and $q_{(j)}\wedge q_{(i)}$ must appear in the sum. 
But conversely for any $j$ with $j+j\equiv k$ the term $q_{(j)}\wedge q_{(j)}$ appears only once.)

Once again we seek actions invariant under the global left action of $G$ and under the local right action of $H$. 
The most general local action with these symmetries is
\be S = -\frac{1}{\lambda^2}\int \sum_{i=1}^{n} \tr \left( \beta_{i} q_{(i)} \wedge \hs q_{(m-i)} 
                                                      + \gamma_{i} q_{(i)} \wedge q_{(m-i)}    \right),\label{action}\ee
for some couplings $\beta_i, \gamma_i, i=1,\dots n$.  

The natural choices of the couplings $\beta_i$ in the kinetic piece are
\be \beta_1 = \beta_2 = \dots = \beta_n = 1 \ee
when $m=2n+1$ is odd, and
\be \beta_1 = \beta_2 = \dots = \beta_{n-1} = 1,\quad \beta_n = \half \ee
when $m=2n$ is even. (The factor $\half$ here is natural because when $m=2n$ there are quadratic terms in 
$q_{(n)}$.) In both the odd and even cases, with these values of the $\beta_i$ the kinetic part of the Lagrangian is simply
\be -\frac{1}{2\lambda^2}\tr (j-A)\wedge \hs (j-A)\label{sigact}\ee
which is nothing but the usual sigma-model Lagrangian on the coset space $G/H$. Henceforth we shall specialise to this
choice. 

(In the case particular case 
\be G/H=\frac{PSU(2,2|4)}{SO(1,4)\times SO(5)},\ee
and more generally whenever $\g$ is a Lie superalgebra and $\sigma$ has order $4$ and respects the bose-fermi grading, in the sense 
that the subspaces of $\g$ of grades 0 and 2 are bosonic while those of grades 1 and 3 are fermionic, then this choice is called the 
``hybrid'' action \cite{BBHZZ} because it includes kinetic terms for both the target space bosons and fermions. The other natural 
choice is the Green-Schwarz action, $\beta_1=0,\beta_2=\half$ \cite{Metsaev:1998it}, which has a kinetic term only for the bosons. 
We will not address the interesting question of how this action should be generalised when $m\neq 4$.)

The motivation for specialising to the kinetic term (\ref{sigact}) here is that it produces the simplest equations of motion, as follows.
Consider the case of $m=2n+1$. To find the equation of motion involving derivatives of $q_{(1)}$ we apply the variation 
$g\mapsto g(1+X)$ with $X$ of grade $-1\equiv 2n$. The kinetic terms vary as follows:
\bea \delta ( q_{(1)} \wedge \hs q_{(2n)} )   &=& \left(q_{(2)} X - X q_{(2)} \right)\wedge \hs q_{(2n)} + q_{(1)} \wedge \hs DX 
\label{kinodd}\\
     \delta ( q_{(2)} \wedge \hs q_{(2n-1)} ) &=& \left(q_{(3)} X - X q_{(3)} \right)\wedge \hs q_{(2n-1)} + q_{(2)} 
                                                                             \wedge \hs \left(q_{(2n)}X - Xq_{(2n)}\right)\nn\\
 &\vdots& \nn\\
  \delta ( q_{(n)} \wedge \hs q_{(n+1)} )  &=& \left(q_{(n+1)} X - Xq_{(n+1)} \right)\wedge \hs q_{(n+1)} + q_{(n)} 
                                                                                \wedge \hs \left(q_{(n+2)}X - Xq_{(n+2)}\right) \nn\eea
and so if one makes the simplest choice and sets $\beta_k=1$ for all $k$ then terms cancel (by cyclicity of the trace in the case of the
$q_{(n+1)}$ term) and one is left with an equation of motion of the form
\be D\hs q_{(1)} = \dots \ee
where the right hand side is the variation of the WZ terms. And, crucially, the same applies to all the other $q_{(i)}$, because the 
choice $\beta_1=\beta_2=\dots=\beta_n=1$ always produces the correct cancellations.\footnote{When $m$ is even the cancellations 
work slightly differently and one is forced to set $\beta_1=\beta_2=\dots=\beta_{n-1}=2\beta_n$, but the argument is similar.}

But consider now the variation of the WZ terms, and again for definiteness 
suppose that $m=2n+1$ is odd. The calculation initially looks very similar, but the crucial difference is that the 
$q_{(n+1)}$ term does not vanish: 
\bea \delta ( q_{(1)} \wedge  q_{(2n)} )   &=& \left(q_{(2)} X - X q_{(2)} \right)\wedge  q_{(2n)} + q_{(1)} \wedge  DX\label{wzodd} \\
     \delta ( q_{(2)} \wedge  q_{(2n-1)} ) &=& \left(q_{(3)} X - X q_{(3)} \right)\wedge  q_{(2n-1)} + q_{(2)} 
                                                                                         \wedge  \left(q_{(2n)}X - Xq_{(2n)}\right) \nn\\
 &\vdots& \nn\\
  \delta ( q_{(n)} \wedge  q_{(n+1)} )  &=& \left(q_{(n+1)} X - Xq_{(n+1)} \right)\wedge  q_{(n+1)} + q_{(n)} 
                                                                                \wedge  \left(q_{(n+2)}X - Xq_{(n+2)}\right) \nn\eea
There are $n$ coefficients $\gamma_k$ to choose and, on making use of the relevant identity in (\ref{Dqk}), $n$ independent terms
on the right-hand sides of these equations. So there is enough freedom to arrange for the total variation from the WZ terms to be 
proportional to $Dq_{(1)}$ and hence for the equation of motion to take the simple form it did in the $\mathbb Z_3$ case,
\be D\hs q_{(1)} \propto D q_{(1)}.\ee
However, the important point is that, for $2n+1 > 3$, it is \emph{not} possible to put the equations of motion for all the $q_{(i)}$ 
in this form simultaneously, because one requires a different choice of the $\gamma_i$ in each case.

This looks rather discouraging at first sight, because in the $\mathbb Z_3$ case the equations of motion (\ref{qeom},\ref{qbareom}) at 
the critical value of the WZ term were particularly simple -- they said that $q$ was covariantly holomorphic while $\qbar$ was 
covariantly anti-holomorphic, and in fact it is known \cite{BBHZZ} that the same is true of $q=q_{(1)}$ and $\qbar=q_{(3)}$ in the 
$\mathbb Z_4$ case -- so one might suspect that the construction of families of flat currents 
somehow relies on this, and that in general the equations of motion have to be
\be D\hs q_{(k)} - D q_{(k)}= 0,\quad D\hs q_{(m-k)} + D q_{(m-k)} = 0\label{wrong}\ee for $k=1,2,\dots n$ when $m=2n+1$ 
(and $k=1,2,\dots n-1$ when $m=2n$; the equation for $q_{(n)}$ takes a different form). 

But in fact what will emerge below is that there \emph{is} a choice of the $\gamma_k$ for which, although the equations 
of motion do not appear to be so elegant, families of flat currents do nevertheless exist, and these currents are the most natural 
generalisation of the those in the $\mathbb Z_3$ and $\mathbb Z_4$ cases. 

To proceed then, we will work backwards by starting with the most obvious ansatz for families of flat currents, and then 
reverse-engineering the correct equations of motion and (the most non-trivial step) the action which produces these equations.

\subsection{Flat currents and the WZ couplings}
Candidate flat currents are of the general form
\be j(\mu) = A + \sum_{i=1}^{m-1} \left( e_i(\mu) q_{(i)} + f_i(\mu) \hs q_{(i)} \right)\label{flatc},\ee
and from the discussion above we know that the equation of motion for each $q_{(k)}$ is of the form 
\be D\hs q_{(k)} + \sum_{(i,j) \in (\mathbb Z_m \setminus \{0\})^2}^{i+j\equiv k} C_k^{ij} q_{(i)} \wedge q_{(j)} =0 \label{Cdef}\ee
for some coefficients $C_k^{ij}=C_k^{ji}$. 
Given these equations of motion, together with the identities (\ref{Fid},\ref{Dqk}), one finds that the current $j(\mu)$ is flat if and 
only if for each $k\in \{1,\dots 2n\}$,
\be e_k e_{m-k} - f_k f_{m-k} = 1, \quad e_k f_{m-k} - e_{m-k} f_k = 0 \label{A}\ee
(these are the conditions at grade 0) and further for all $i,j$ such that $i+j\equiv k$,
\be e_i f_j - f_i e_j = 0, \quad e_i e_j - f_i f_j = e_{k} + C_k^{ij} f_{k}\ee
(these are the conditions at grade $k$).

One can usefully re-express these conditions as follows: For each $k=1,\dots,m-1$ 
\be (e_k + f_k)(e_{m-k} - f_{m-k}) = 1,\label{c0}\ee
and further for all $i,j$ such that $i+j\equiv k$
\be (e_i + f_i)(e_j - f_j) = e_k + C_k^{ij} f_k\label{ck}.\ee

Let us try to solve these equations by setting 
\be e_k + f_k = \mu^k, \quad e_k - f_k = \frac{1}{\mu^{m-k}} \label{para}\ee
with $\mu$ the sole remaining free parameter. This is the natural generalisation of the solution (\ref{para3}) in the 
$\mathbb Z_3$ case, and it has the merit of automatically satisfying (\ref{c0}) and also 
$(e_i + f_i)(e_j - f_j) = (e_i - f_i)(e_j + f_j)$, which is necessary if (\ref{ck}) is to hold.

Given (\ref{para}), the coefficients $C_k^{ij}$ are uniquely determined by (\ref{ck}): we must have
\be C_k^{ij} = \begin{cases} +1 & i+j>m,\\ -1 & i+j<m.\end{cases}\label{Ck}\ee
(Here the relations $\gtrless$ refer to the ordering of $\mathbb Z$. So for example when $m=5$, $C_3^{12}=-1$ since
$1+2<5$ but $C_3^{44}=+1$ since $4+4 (\equiv 3 \mod 5) = 8 > 5$.) 

Thus it is certainly possible to choose equations of motion such that a one-parameter family of flat currents exists -- this is not so 
surprising in itself. What is not at all obvious is that these equations of motion may be obtained from an action of the form 
(\ref{action}). When $m=2n+1$ there are only $n$ free real parameters $\gamma_k$ to choose, and it is necessary to get the values of 
$2n^2$ coefficients $C_k^{ij}$ correct;
when $m$ is even the counting is slightly modified but the apparent problem is the same.

However, it turns out there is a solution. The values
\be \gamma_k =  1 - \frac{2k}{m} \label{gammas}\ee
for the WZ couplings \emph{do} produce the equations of correct equations of motion. The calculation is straightforward but lengthy,
so we shall only sketch it. One computes all the coefficients $C_k^{ij}$ in the equations of motion (\ref{Cdef}) by varying the action
(\ref{action}). There are six cases to consider, which occur naturally in pairs
\bea k\leq n,\, i,j>n;  &&  k>n,\, i,j\leq n; \\
     k\leq n,\, i,j<k;  &&  k>n,\, i,j > k;\\     
     k<i\leq n \text{ (and } j>n); &&  k>i>n \text{ (and } i\leq n).\eea       
When one demands that the $C_k^{ij}$ take the values given in (\ref{Ck}) the first pair of cases both produce the same condition:
\be \text{ for all } i, j\leq n \text{ such that } i+j>n,\quad    \gamma_i + \gamma_j + \gamma_{2n+1-i-j} = +1,\ee
while the final four cases all separately produce the condition
\be \text{ for all } i, j\leq n,\text{ such that } i+j\leq n,\quad\gamma_i + \gamma_j - \gamma_{i+j} = + 1.\ee
Naive counting would suggest that these conditions still constitute an over-determined set of equations for the $\gamma_i$, 
but nevertheless they are satisfied by (\ref{gammas}). 

Our result is thus that, for this choice of couplings $\gamma_i$, there exists a one-parameter family of flat currents, of the
form (\ref{flatc}) with the coefficient functions 
\be e_{(k)} = \frac{\mu^{m}+1}{2\mu^{m-k}},\quad f_{(k)} = \frac{\mu^{m}-1}{2\mu^{m-k}}.\ee

As in the $\mathbb Z_3$ case, the trivial solution is $\mu=1$ and we expand around this by setting $\mu=\exp \theta$ and find, 
to first order in $\theta$, 
\be e_{(k)} = 1 - \half(m-2k)\theta+\dots,\quad f_{(k)} = \half m\theta+\dots.\ee
so that once again it is possible to expand the corresponding gauge-invariant current 
$g(j(\mu)-j)g^{-1}$
and find at first order (a multiple of the hodge dual of) the Noether current $J$ of the global $G$ symmetry, which is
\be J = \sum_{k=1}^{m-1} g\left( q_{(k)} -\left(1-\frac{2k}{m}\right) \hs q_{(k)} \right)g^{-1}.\ee
As an additional check, in the particular case of automorphisms of order $m=4$, the flat currents found here coincide with those 
found in \cite{Vallilo:2003nx}. (To connect the notations: the $(A,\bar A)$ of \cite{Vallilo:2003nx} is here $j(\mu)-j$, and this 
accounts for the many subtractions of one which occur in the parameterisation (3.7) of the flat currents in that paper. 
The solutions are then identical, with the spectral parameters related by $\mu_\text{here}= \mu_\text{there}^{-\half}$.)

It is worth noting that the argument above also shows that the equations of motion 
(\ref{wrong}) that one might naively prefer to have, but which we showed were incompatible with actions of the form (\ref{action}),
are not compatible with the existence of flat currents either, at least in any obvious way. 

We conclude by remarking that the values (\ref{gammas}) of the WZ couplings make the WZ 3-form look particularly simple: 
one finds
\be d\sum_{k=1}^n \left(1-\frac{2k}{m} \right) \tr q_{(k)}\wedge \hs q_{(m-k)} = 
 \sum_{\{i,j,k\}\subset \mathbb Z}^{i+j+k=m} \tr\left( q_{(i)} \wedge q_{(j)} \wedge q_{(k)} 
                                                      -q_{(m-i)}\wedge q_{(m-j)}\wedge q_{(m-k)} \right),\ee
and it is only for this choice of the 2-form on the left that all the coefficients of the traces in the sum on the right are unity.

\vspace{0.5cm}

\emph{Acknowledgements } I thank Jonathan Evans and Niall Mackay for reading through earlier drafts of this work, and
David Kagan and Aninda Sinha for useful discussions. I gratefully acknowledge the financial support of PPARC.

\end{document}